\documentclass[12pt]{article}
\usepackage[T1]{fontenc}
\usepackage[utf8]{inputenc}
\usepackage[english]{babel}
\usepackage{amsmath,amssymb}
\usepackage[a4paper,margin=25mm,nohead,footskip=10mm]{geometry}
\usepackage[hidelinks]{hyperref}

\title{On a model of quantum quadratic gravity in $7+1$-dimensional spacetime}
\author{%
    V.~Chistiakov\\[0.3em]
    \small Lomonosov Moscow State University\\
    \small \href{mailto:vsevolod.chistyakov@gmail.com}{\nolinkurl{vsevolod.chistyakov@gmail.com}}
    \and
    A.~Mingaleev\\[0.3em]
    \small Lomonosov Moscow State University\\
    \small \href{mailto:mingaleevartur869@gmail.com}{\nolinkurl{mingaleevartur869@gmail.com}}
}
\date{}

\begin{document}
\maketitle

\begin{abstract}
We consider a quadratic gravity model supplemented by the terms $\nabla_{\mu}R\nabla^{\mu}R$ and $(\nabla^{2}R)^2$. For specific relations among the coupling constants, the action takes a perfect-square form. In the spatially flat FLRW model, we perform a reduction to a dynamical variable invariant under time reparameterizations. We construct a nonlinear change of variable that transforms the reduced action into the Gaussian action associated with the Wiener measure. An analysis of the transformation Jacobian within a time discretization singles out the case of $7+1$-dimensional spacetime, in which the Jacobian of the discrete transformation under consideration is independent of the dynamical variable. On this basis, we propose a formal representation of expectation values of observables in the reduced model as integrals with respect to the Wiener measure.
\end{abstract}

\section{Introduction}
Quadratic gravity extends the Einstein--Hilbert action by terms quadratic in the curvature. In four dimensions, the general theory of this type is perturbatively renormalizable owing to the improved ultraviolet behavior of the propagator \cite{Stelle1977}. Curvature-squared corrections also play an important role in inflationary scenarios \cite{Starobinsky1980,Rub} and quantum cosmology \cite{Slav,R+R2}. Buchbinder and Lyakhovich~\cite{Buhb} carried out the canonical quantization of $R^2$ gravity and constructed a local path integral measure.

An approach to constructing path integral measures for quadratic gravity in the FLRW metric was proposed in Refs.~\cite{2QG,3QG}. The integration is performed over a gauge-invariant dynamical variable $G(\tau)$. Nonlinear, nonlocal changes of variables map the path integral measure of quadratic gravity to the Wiener measure. This approach was used to compute the first nontrivial quantum corrections to the classical scale factor.

In this work, we consider a model with additional terms $\nabla_{\mu}R\nabla^{\mu}R$ and $(\nabla^{2}R)^2$, characteristic of higher-derivative gravity \cite{Asorey1997}. For specific relations among the coupling constants, the action takes a perfect-square form. We construct a nonlinear change of variable $G=\Phi(p)$. In $7+1$ dimensions, we obtain a representation of expectation values of observables as integrals with respect to the Wiener measure.

We first summarize the main results of Refs.~\cite{2QG,3QG}. In an arbitrary gauge, the metric of a homogeneous and isotropic universe is parametrized by two functions $(N(t),a(t))$:
\begin{equation}\label{FLRW}
    ds^2=N^2(t)dt^2-a^2(t)d{\bf x}^2\, .
\end{equation}
The action, regarded as a functional of these variables, takes the form
\begin{equation}\label{actNa}
    A(N,a)=\int dt \sqrt{|g|} \left(\Lambda +\kappa R(N,a) +\beta R^2(N,a)\right)\, .
\end{equation}
It is invariant under the group $\mathrm{Diff}^1_{+}(\mathbb{R}^{+})$ of reparameterizations of the time half-line. Transformations $\phi\in \mathrm{Diff}^1_{+}(\mathbb{R}^{+})$ act as follows:
\begin{equation}\label{Nadiff}
    (\phi N)(t)= \phi'(t)N(\phi(t))\,, \qquad (\phi a)(t)=a(\phi(t))\, .
\end{equation}
The expectation value of an observable $F(N,a)$ is expressed as a functional integral over the space of variables $(N(t),a(t))$:
\begin{equation}
    \langle F \rangle=\frac{\int F(N,a)\exp\left(-A(N,a)\right)\mathcal{M}(dNda)}{\int \exp\left(-A(N,a)\right)\mathcal{M}(dNda)}\, .
\end{equation}
The measure $\mathcal{M}(dNda)$ is assumed to be invariant under transformations (\ref{Nadiff}).
In Refs.~\cite{2QG,3QG}, a transformation from the variables $(N(t),a(t))$ to the variables $(G(\tau),\varphi(\tau))$ was constructed such that
\begin{equation}
    (\phi G)(\tau)=G(\tau)\, ,\qquad (\phi\varphi)(\tau)=(\varphi\circ\phi)(\tau)\equiv\varphi(\phi(\tau))\, .
\end{equation}
In the coordinates $(G(\tau),\varphi(\tau))$, the measure $\mathcal{M}(dNda)$ is a product measure \cite{2QG,3QG}:
\begin{equation}
    \mathcal{M}(dNda)=\mu(dG)\chi(d\varphi)\, ,
\end{equation}
where $\chi(d\varphi)$ is right-invariant:
\begin{equation}
    \chi(d(\varphi\circ\phi))=\chi(d\varphi)\, .
\end{equation}
This factorization allows the expectation value of an observable to be written as a functional integral over the space of invariants $G(\tau)$:
\begin{equation}\label{Faver}
    \langle F\rangle =\frac{\int F(N,a)\exp\left(-A(N,a)\right)\mathcal{M}(dNda)}{\int \exp\left(-A(N,a)\right) \mathcal{M}(dNda)}=\frac{\int F(G) \exp\left(-A(G)\right)\mu(dG)}{\int \exp\left(-A(G)\right)\mu(dG)}\, .
\end{equation}
In Refs.~\cite{2QG,3QG}, a change of variable $G=\Phi(p)$ was constructed that expresses expectation values of observables as integrals with respect to the Wiener measure:
\begin{equation}
    \int F(G)\exp\left(-A(G)\right)\mu(dG)=\int F(\Phi(p))w(dp)\, .
\end{equation}

\section{Quadratic gravity with higher-derivative terms}
Consider a quadratic gravity model supplemented by covariant derivatives of the curvature:
\begin{equation}\label{main action}
\mathcal{A}=\int d^{D+1}x\sqrt{|g|} \left(\Lambda -\frac{\varkappa}{2D} R+\frac{\lambda^2}{8D^2}R^2+\alpha \nabla_{\mu}R\nabla^{\mu}R+\frac{1}{8D^2\sigma^2}\left(\nabla^{2}R\right)^2 \right)\, .
\end{equation}
If the model parameters satisfy the relations
\begin{equation}
\Lambda=\frac{\varkappa^2}{2\lambda^2}\, ,\qquad \alpha=\frac{\lambda}{4D^2 \sigma}\, ,
\end{equation}
the action takes the form
\begin{equation}
\mathcal{A}=\frac{1}{2\sigma^2}\int d^{D+1}x\sqrt{|g|}\left(\frac{\varkappa\sigma}{\lambda} - \lambda \sigma \frac{R}{2D}+\left(\nabla^{2}\frac{R}{2D}\right)\right)^2\, .
\end{equation}
If the parameters also satisfy the relation
\begin{equation}
    \lambda=-\left(\frac{4D^2}{D-1}\right)^{1/3} \left(\frac{\varkappa}{\sigma}\right)^{1/3}\, ,
\end{equation}
the model admits a classical solution of the form
\begin{equation}
    a(t)=a(0)\exp\left(-\frac{\epsilon}{4D}t^2\right)\, ,\qquad \epsilon\equiv\left(\frac{4D^2}{D-1}\right)^{1/3}(\varkappa\sigma^2)^{1/3}\, .
\end{equation}
In this case, the action takes the form
\begin{equation}\label{fullA}
\mathcal{A}=\frac{1}{2\sigma^2}\int d^{D+1} x\sqrt{|g|}\left(-\frac{D-1}{4D^2}\epsilon^2+\epsilon \,\frac{R}{2D}+\left(\nabla^{2}\frac{R}{2D}\right)\right)^2\, .
 \end{equation}
We evaluate the action (\ref{fullA}) in the metric
\begin{equation}\label{GFLRW}
ds^2=(G'(\tau))^2(d\tau^2-d{\bf x}^2)\, .
\end{equation}
Since the integrand in (\ref{fullA}) is independent of the spatial coordinates ${\bf x}$, following Refs.~\cite{2QG,3QG}, we define
\begin{equation}
    A(G)=\frac{\mathcal{A}(G)}{\int d^{D}{\bf x}}\, .
\end{equation}

The reduced action $A(G)$ takes the form
\begin{equation}
    A(G)=\frac{1}{2\sigma^2}\int_0^T\, d\tau \, \,\, (\mathcal{F}(G))^2\, ,
\end{equation}
where
\begin{equation}
   \mathcal{F}(G)=(G')^{\frac{D-9}{2}}G^{(5)}+2(D-5)(G')^{\frac{D-11}{2}}G''G^{(4)}+(D-6)(G')^{\frac{D-11}{2}}(G''')^2+
\end{equation}
\begin{equation*}
    +(D^2-17D+48)(G')^{\frac{D-13}{2}}(G'')^{2}G'''-2(D-3)(D-6)(G')^{\frac{D-15}{2}}(G'')^4+
\end{equation*}
\begin{equation}
    +\epsilon\left((G')^{\frac{D-5}{2}}G'''+\frac{D-3}{2}(G')^{\frac{D-7}{2}}(G'')^2\right)+(G')^{\frac{D+1}{2}}\frac{D-1}{4D^2}\epsilon^2\, .
\end{equation}

\section{Transformation of the path integral measure to the Wiener measure}

We construct a change of the dynamical variable, $G=\Phi(p)$, such that
\begin{equation}
A(\Phi(p))=\frac12\int^{T}_{0} d\tau(p'(\tau))^2\, .
\end{equation}
To this end, it suffices to require that $G(\tau)$ satisfy the equation
\begin{equation}\label{Geq}
    \mathcal{F}(G)=\sigma p'\, 
\end{equation}
subject to the initial conditions
\begin{equation}
    G(0)=0\, ,\qquad G'(0)=1\, ,\qquad G''(0)=0\,, \qquad G'''(0)=-\frac{\epsilon}{2D}\, ,\qquad G^{(4)}(0)=0\, .
\end{equation}

For $D=7$, the expectation value of the observable $F(G)$ in (\ref{Faver}) can be expressed as an integral with respect to the Wiener measure:
\begin{equation}\label{FGw}
    \langle F\rangle =\int F(\Phi(p))w(dp)\, .
\end{equation}

To establish (\ref{FGw}), we introduce the auxiliary function
\begin{equation}\label{Gbeta}
\beta=\frac{G''}{G'}\, ,\qquad G'=\exp\left(\int_{0}^{\tau}\beta(s) ds\right)\, .
\end{equation}
The following identity holds \cite{3QG}:
\begin{equation}\label{Gbetaint}
    \frac{\int F(G)\exp\left(-A(G)\right)\mu(dG)}{\int \exp\left(-A(G)\right)\mu(dG)}=\frac{\int F(G(\beta))\exp\left(-A(G(\beta))\right)d\beta}{\int\exp\left(-A(G(\beta))\right)d\beta}\, .
\end{equation}

Using (\ref{Geq}) and (\ref{Gbeta}), we write the transformation from $\beta(\tau)$ to $p(\tau)$ as
\begin{equation}\label{pbetaX}
\sigma p(\tau) =\int_0^{\tau} ds\,\mathcal{X}(\beta(s))\, , 
\end{equation}
where
\begin{equation}
    \mathcal{X}(\beta)\equiv \mathcal{F}(G(\beta))=\exp\left(\frac{D-7}{2}\int_0^{\tau}\beta(s)\,ds\right)\times
\end{equation}
\begin{equation*}
     \times\left\{ \beta''' + \beta^{2} \beta' \left( D^{2} - 9 D + 12\right) + (2 \beta \beta'' + \beta'^{2}) \left(D-3\right) -\beta^{4} (D-1)(D-3) \right\}+
\end{equation*}
\begin{equation*}
    +\frac{\epsilon}{2}\exp\left(\frac{D-3}{2}\int_0^{\tau}\beta(s)\, ds\right) \left(\beta^{2} \left(D - 1\right) + 2 \beta'\right)+
\end{equation*}
\begin{equation*}
    +\frac{D-1}{4D^2}\epsilon^2 \exp\left(\frac{D+1}{2}\int_0^{\tau}\, \beta(s)\, ds\right)\, .
\end{equation*}

For $D=7$, we rewrite (\ref{pbetaX}) as
\begin{equation}\label{pbeta}
    \sigma p(\tau)=\beta''(\tau)+\int_0^{\tau}\, h\left(\beta''(s),\beta'(s),\beta(s),\int_0^{s} \beta(s_1)ds_1\right)\, ds\, ,
\end{equation}
where $h$ is a continuous function whose explicit form is not needed here. Assuming that $\beta''(\tau)$ is continuous, we discretize this equation. We partition the interval $[0,\tau]$ into $N$ subintervals with endpoints $\tau_0=0,\tau_1,...,\tau_N=\tau$. Set $\Delta=\tau_{i+1}-\tau_i$, $\beta_i=\beta(\tau_i)$, $p_{i}=p(\tau_i)$. Interpreting the integral in the It\^o sense (with the integrand evaluated at the left endpoint of each subinterval), we rewrite Eq.~(\ref{pbeta}) as
\begin{equation}
     p_N=\frac{\beta_N-2\beta_{N-1}+\beta_{N-2}}{\sigma \Delta^2} + \tilde h(\beta_{N-1},\beta_{N-2},...,\beta_0)\, .
\end{equation}
The Jacobian matrix
$$\frac{\partial p_i}{\partial \beta_j}$$
is therefore lower triangular, and its determinant is
\begin{equation}
    \det \frac{\partial p_i}{\partial \beta_j}=(\sigma\Delta^2)^{-N}\, .
\end{equation}
This yields the equality of measures
\begin{equation}\label{betapint}
    \frac{\int F(G(\beta))\exp\left(-A(G(\beta))\right)d\beta}{\int\exp\left(-A(G(\beta))\right)d\beta}=\int F(G(\beta(p)))w(dp)\, .
\end{equation}
Combining (\ref{Gbetaint}) and (\ref{betapint}) gives (\ref{FGw}).

\section{Conclusions}
In this paper, we have shown that the approach developed in Refs.~\cite{2QG,3QG} to constructing a functional measure in quadratic gravity can be extended to a broader class of theories of the form (\ref{main action}). In $7+1$-dimensional spacetime, the functional measure on the space of invariants $G(\tau)$ for the model (\ref{fullA}) with the FLRW metric (\ref{GFLRW}) reduces to the Wiener measure.

The established properties of functional measures in theories of the form (\ref{main action}) will enable the computation of expectation values of observables, such as the scale factor $a(t)$, by evaluating integrals with respect to the Wiener measure. These calculations will be presented in a subsequent paper.

\vspace{1cm}
\noindent The work of V. Chistiakov is supported by the "Basis" Foundation.


\begin{thebibliography}{99}

\bibitem{Stelle1977}
K.~S.~Stelle, \emph{Renormalization of higher-derivative quantum gravity},
Physical Review D \textbf{16}(4), 953--969 (1977).
\href{https://doi.org/10.1103/PhysRevD.16.953}{doi: 10.1103/PhysRevD.16.953}.

\bibitem{Starobinsky1980}
A.~A.~Starobinsky, \emph{A new type of isotropic cosmological models without singularity},
Physics Letters B \textbf{91}(1), 99--102 (1980).
\href{https://doi.org/10.1016/0370-2693(80)90670-X}{doi: 10.1016/0370-2693(80)90670-X}.

\bibitem{Rub}
D.~S.~Gorbunov, V.~A.~Rubakov,
\emph{Introduction to the Theory of the Early Universe: Cosmological Perturbations and Inflationary Theory},
World Scientific, Singapore (2011).
\href{https://doi.org/10.1142/7873}{doi: 10.1142/7873}.

\bibitem{Slav}
A.~Salvio, \emph{Quadratic Gravity},
Frontiers in Physics \textbf{6}, 77 (2018).
\href{https://doi.org/10.3389/fphy.2018.00077}{doi: 10.3389/fphy.2018.00077}.

\bibitem{R+R2}
A.~K.~Sanyal, B.~Modak, \emph{Quantum cosmology with $R+R^2$ gravity},
Classical and Quantum Gravity \textbf{19}(3), 515--525 (2002).
\href{https://doi.org/10.1088/0264-9381/19/3/307}{doi: 10.1088/0264-9381/19/3/307}.

\bibitem{Buhb}
I.~L.~Buchbinder, S.~L.~Lyakhovich,
\emph{Canonical quantization and local measure of $R^2$ gravity},
Classical and Quantum Gravity \textbf{4}(6), 1487--1501 (1987).
\href{https://doi.org/10.1088/0264-9381/4/6/008}{doi: 10.1088/0264-9381/4/6/008}.

\bibitem{2QG}
V.~V.~Belokurov, E.~T.~Shavgulidze, \emph{Path integrals in quadratic gravity},
Journal of High Energy Physics \textbf{2022}(02), 112 (2022).
\href{https://doi.org/10.1007/JHEP02(2022)112}{doi: 10.1007/JHEP02(2022)112}.

\bibitem{3QG}
V.~V.~Belokurov, V.~V.~Chistiakov, E.~T.~Shavgulidze,
\emph{Perturbation theory for path integrals in quadratic gravity},
International Journal of Modern Physics A \textbf{40}(33), 2540005 (2025).
\href{https://doi.org/10.1142/S0217751X25400056}{doi: 10.1142/S0217751X25400056}.

\bibitem{Asorey1997}
M.~Asorey, J.~L.~L\'opez, I.~L.~Shapiro,
\emph{Some remarks on high derivative quantum gravity},
International Journal of Modern Physics A \textbf{12}, 5711--5734 (1997).
\href{https://doi.org/10.1142/S0217751X97002991}{doi: 10.1142/S0217751X97002991};
\href{https://arxiv.org/abs/hep-th/9610006}{arXiv:hep-th/9610006}.

\end{thebibliography}
\end{document}